\documentclass[iop]{emulateapj}
\usepackage{graphicx}
\usepackage{subfigure}
\usepackage{epsfig}

\begin{document}

\title{Cosmic Transients Test Einstein's Equivalence Principle out to GeV Energies}

\author{He Gao$^{1,2,3,4,6}$, Xue-Feng Wu$^{1,2,5}$, Peter M\'esz\'aros$^{2,3,4}$}

\affil{$^1$Purple Mountain
Observatory, Chinese Academy of Sciences, Nanjing 210008, China; xfwu@pmo.ac.cn
\\$^2$Department of Astronomy and Astrophysics, Pennsylvania State University, 525 Davey Laboratory, 
University Park, PA 16802: \\
$^3$Department of Physics, Pennsylvania State University, 525 Davey Laboratory, University Park, PA 
16802\\
$^4$Center for Particle and Gravitational Astrophysics, Institute for Gravitation and the Cosmos, 
Pennsylvania State University, 525 Davey Laboratory, University Park, PA 16802\\
$^5$Joint Center for Nuclear Particle Physics and Cosmology, Purple Mountain Observatory -
Nanjing University, Nanjing 210008, China\\
$^6$Department of Astronomy, Beijing Normal University, Beijing 100875, China}

\begin{abstract}
The Einstein Equivalence Principle (EEP) can be probed with astrophysical sources emitting
simultaneously different types of neutral particles, or particles with varying 
energies, by testing their time of flight through the same gravitational field.
Here we use the time delays between correlated photons from cosmological 
transients to constrain the accuracy of the EEP. 
We take data from two gamma-ray bursts as an example, and use, as a lower limit to the theoretical time delays between different energies, delays arising from only the gravitational field of our own galaxy. We then show that the parameterized post-Newtonian parameter $\gamma$ is the same for photons over energy ranges between eV and MeV and between MeV and GeV to a part in $10^{-7}$, which is at least one order of magnitude better than previous limits. Combining this bound on the wavelength dependence of $\gamma$ with the absolute bound $|\gamma-1|<0.3\%$ from light-deflection measurements at optical (eV) wavelengths, we thus extend this absolute bound on $\gamma$ to GeV energies.
\end{abstract}

\section{INTRODUCTION}

One statement of the Einstein Equivalence Principle (EEP) is that any uncharged test body traveling in 
empty space will follow a trajectory independent of its internal structure and composition. For test 
bodies with macroscale masses, the accuracy of the EEP can be measured in a Newtonian context. 
However, for test particles like photons and neutrinos, Newtonian dynamics is no longer precise 
enough. To test the accuracy of the EEP at this level, at least the lowest-order deviations from 
Newton's law of universal gravitation are required, which can be conveniently described by the 
parameterized post-Newtonian (PPN) formalism. For all metric theories of gravitation in which all 
bodies satisfy the EEP, the PPN formalism explicitly details the parameters in which a general theory of 
gravity can differ from Newtonian gravity. Each theory is specified by a set of numerical coefficients 
(PPN parameters) see \cite[][for a review]{will06,will14}. 

A gravity theory can be tested through the absolute values of PPN parameters. For example, it has 
been pointed out that the time interval required for photons to traverse a given distance is longer in 
the presence of a gravitational potential $U(r)$ by 
\begin{eqnarray}
\delta t=-\frac{1+\gamma}{c^3}\int_{r_e}^{r_o}~U(r)dr
\label{eq:gra}
\end{eqnarray}
where $r_e$ denotes the location of the source of particle emission and $r_o$ that of observation 
\citep{shapiro64,krauss88,longo88}. Half of the delay is caused by the warping of space by the 
gravitational field, the other half can be thought of as a result of gravitational redshift. Here $
\gamma$ is one of the PPN parameter, representing how much space curvature is produced by unit 
rest mass. Different gravity theories predict different values for $\gamma$. For instance, general 
relativity predicts that $\gamma=1$ \citep{will93}. Many precise methods have been devised to 
test the accuracy of general relativity through the measurement of the value of $\gamma$, 
the most precise being the measurement of the gravitational deflection of light near the Sun and 
the round-trip travel time delay of an artificial radar signal due to the solar system gravity 
\citep{will06,will14}:
\begin{itemize}
\item the most precise results from the deflection of light is through very-long-baseline radio interferometry measurement, yielding $\gamma-1= (-0.8 \pm 1.2) \times 10^{-4}$~\citep{lambert09,lambert11};
\item the most precise results from the travel time delay of a radar signal, which is also the most precise constraints on the $\gamma$ value to date, is from the Doppler tracking of the Cassini spacecraft, yielding $\gamma -1 = (2.1 \pm 2.3) \times 10^{-5}$ \citep{bertotti03}. 
\end{itemize}

It is worth pointing out that $\gamma=1$ is not a sufficient condition to identify general relativity, 
since it is not the only theory that predicts $\gamma=1$ \citep{will93}. Regardless of the absolute 
value of $\gamma$, all metric theories of gravity that incorporate the EEP predict that all test 
particles (photons or neutrinos) must follow identical trajectories and undergo the same time delay, 
even though the delay time may vary between different theories. In other words, as long as the EEP is 
valid, all metric theories predict $\gamma=\gamma_{\rm 1}=\gamma_{\rm 2}$ (and $\gamma$ need  not
even be 1), where ${\rm 1}$ and ${\rm 2}$ represent two different test particles. In this case, the 
accuracy of the EEP can be characterized by limits on the differences of $\gamma$ value for 
different species of particles, or the same specie of particle with varying energies. 

It has been proposed that the close coincidence in time of arrival of the photon and neutrino bursts 
from the supernova 1987A provides a strong test of the EEP  \citep{longo88,krauss88}. 
Comparing the arrival time delay of the photons and neutrinos, \cite{longo88} proposed that the $
\gamma$ values for photon (eV) and neutrino (MeV) are identical to an accuracy of approximately 
$0.37\%$. On the other hand, the very small dispersion in arrival times observed for neutrinos 
of different energies ($7.5-40~{\rm MeV}$) can be used to set a limit on the energy dependence of $
\gamma_{\nu}$ to the level of $1.6\times 10^{-6}$.  

Supernova 1987A is an extragalactic source, but non-cosmological. In this work, we propose that 
the time delays between correlated photons from cosmological transients, such as Gamma-Ray Bursts (GRBs), 
can also be used to constrain the EEP with similar approaches. In the following, we first discuss the 
general picture of using cosmic transients to constrain the EEP. Taking GRBs as an example, we then 
show the ability of this method for improving the constraints, and discuss the uncertainties involved.

\section{General description for deriving EEP constraints with cosmic transients}

Cosmic transient sources often show some time delays between different energy bands. The 
main terms that might contribute to this time delay may be expressed as follows:
\begin{eqnarray}
\Delta t_{\rm obs}=&&\Delta t_{\rm int}+\Delta t_{\rm LIV}+\Delta t_{\rm spe}+\Delta t_{\rm gra}
\label{eq:deltat}
\end{eqnarray}
In the following, we discuss each component in turn.
\begin{itemize}
\item $\Delta t_{\rm int}$ is the intrinsic  (astrophysical) time delay between two test photons, 
and this has the largest 
uncertainty since it is impossible to determine it based on the observational data alone. Appropriate 
assumptions for the radiation mechanism are usually required to assign a chronological order for the
different energy photons, but it is still hard to estimate the exact value of $\Delta t_{\rm int}$. 
\item $\Delta t_{\rm LIV}$ represents the potential time delay due to Lorentz invariance violation (LIV)
via an energy-dependent velocity of light. Some descriptions of it are based on quantum fluctuations in 
the back- ground space-time metric causing a non-trivial dispersion in vacuo for light propagating 
through this space-time foam \citep{amelino98}. Taking into account the effect on the propagation of 
photons due to the expansion of the Universe, even a tiny variation in photon speed may lead to an 
observable  time delay \citep{ellis06,abdo09},
\begin{equation}
\Delta t_{LIV} =\frac{(1+n)}{2 H_0} \frac{E_1^n-E_2^n}{E_{QG,n}^n}  \int_0^z  \frac{(1+z')^ndz'}{h(z')},
\label{eq:timedelay}
\end{equation}
which is the n-th (first non-vanishing term) of a series expansion in an effective field theory
formulation of quantum gravity including Lorentz invariance violating terms. Here
$E_1$ and $E_2$ represent the energies for two test photons ($E_1>E_2$), $E_{QG}$ is the 
effective energy scale of quantum gravity, $h(z)=H(z)/H_0$, with $H_0$ being the present Hubble expansion 
rate and $H(z)$ the Hubble parameter at different redshifts. For a limiting $E_{QG}$ equal
to the Planck energy, the recent evidence disfavors the existence of linear (order $n=1$) LIV terms, i.e. energy-dependent departures of the speed of photons from the
usual speed of light \citep{ellis06,abdo09,pan15}. Substituting the Plank energy for $E_{QG}$ it 
is easy to show that the higher order terms of the expansion ($n>2$), even if they exist, are 
negligible for the purposes of this work \footnote{In general, one can find that as long as $E_{QG,n}/E_1\gg10^{18/n}$, the $n$-th expansion of $\Delta t_{LIV}$ becomes negligible for the purposes of this work (e.g., much smaller than the order of a second).}. We thus ignore this term in the following analysis.
\item $\Delta t_{\rm spe}$ represents the potential time delay due to special-relativistic effects
in the case where the photons have a rest mass which is non-zero, which reads
\begin{eqnarray}
\Delta t_{\rm spe}= \frac{m_{\rm ph}^2c^4}{2H_0}\left(\frac{1}{E_1^2}-\frac{1}{E_2^2}\right)  
\int_0^z  \frac{(1+z')dz'}{h(z')},
\end{eqnarray}
where $m_{\rm ph}$ is the photon rest mass. Modern experiments have provided the upper 
limits for the photon rest mass as $m_{\rm ph}<10^{-18}~{\rm eV/c^2}$ \citep{amsler08}. In this 
case, $\Delta t_{\rm spe}$ is negligible even when the energy of test photons are lower than radio 
band. 
\item $\Delta t_{\rm gra}$ corresponds to the difference in arrival time of two photons of energy 
$E_1$ and $E_2$, caused by the gravitational potential $U(r)$ integrated from the emission site to 
the Earth, which can be estimated with equation \ref{eq:gra}, as long as the expression of $U(r)$ is known,
\begin{eqnarray}
\Delta t_{\rm gra}=\frac{\gamma_{\rm 1}-\gamma_{\rm 2}}{c^3}\int_{r_o}^{r_e}~U(r)dr
\label{eq:loc}
\end{eqnarray}
where $c=3\times10^{10}~{\rm cm~s^{-1}}$ is the energy-independent speed for massless particles. 
In general, $U(r)$ can be divided into three parts: the gravitational potential of our 
galaxy $U_{\rm mw}(r)$ would be dominant out to a certain distance, say to the edge of the local 
group ($\sim1$ Mpc). Based on the cosmological principle and current CMB results, the gravitational 
potential should gradually approach a flat background intergalactic potential $U_{\rm IG}(r)$ between 
our galaxy and the host galaxy (or cluster) of the transient. This assumption may be valid at least for 
sources with $z<2$, for which the lensing probability is significant only to $< 1\% $ \citep{wiklind13}. 
The final contribution is from the gravity well of the host galaxy (or the relevant galaxy cluster) 
$U_{\rm host}(r)$. 
\end{itemize}
Leaving out the negligible components, equation \ref{eq:deltat} is
\begin{eqnarray}
\Delta t_{\rm obs}=\Delta t_{\rm int}+\Delta t_{\rm gra}
\end{eqnarray}
For a specific case where high energy photons arrive later than the low energy photons 
($\Delta t_{\rm obs}>0$), if we know that $\Delta t_{\rm int}>0$, we will have 
\begin{eqnarray}
\Delta t_{\rm obs}>\frac{\gamma_{\rm 1}-\gamma_{\rm 2}}{c^3}\int_{r_o}^{r_e}~U(r)dr
\label{eq:deltatnew}
\end{eqnarray}
The potential function for $U_{\rm IG}(r)$ and $U_{\rm host}(r)$ is hard to model, but it is 
very likely that the effect of these two terms is much larger than if we simply assumed that the
potential is just $U_{\rm mw}(r)$ extended to the distance of the transient, i.e., 
\begin{eqnarray}
\Delta t_{\rm obs}>\frac{\gamma_{\rm 1}-\gamma_{\rm 2}}{c^3}\int_{r_o}^{r_e}~U_{\rm mw}(r)dr
\label{eq:deltatnew}
\end{eqnarray}
Although the gravitational potential of the Milky Way at large distances is still not well known, 
\cite{krauss88} examined two popular potential models, e.g., the Keplerian potential $U(r)=-GM/r$ 
and the isothermal potential $U(r)=v_c^2[\ln (r/r_{\rm max})-1]$, for $r<r_{\rm max}$, where $v_c$ 
is the circular speed at the solar radius and the potential becomes Keplerian for $r>r_{\rm max}
=100~{\rm kpc}$. They proposed that different models for $U_{\rm mw}(r)$ do not have a strong 
influence on the results for testing EEP. In this work, we adopt the Keplerian potential for our galaxy.
Thus, for the purposes of obtaining a lower limit, we can simply extend this galactic potential 
out to cosmic scales to bracket from below the potential function of $U_{\rm IG}(r)$ and 
$U_{\rm host}(r)$. We thus have 
\begin{eqnarray}
\gamma_{\rm 1}-\gamma_{\rm 2}<\Delta t_{\rm obs}\left(\frac{GM_{\rm mw}}{c^3}\right)^{-1}
\ln^{-1}\left(\frac{d}{b}\right)
\label{eq:gamma}
\end{eqnarray}
where $G=6.68\times10^{-8}~{\rm erg~cm~g^{-2}}$ is the gravitational constant, $M_{\rm mw}
\approx6\times10^{11}M_{\odot}$ is the mass of our galaxy \citep{mcmillan11,kafle12}, $d$ is the 
distance from the transient to Earth, and $b$ is the impact parameter of the light rays relative to 
the center of the galaxy. For a cosmic source in the direction (${\rm R.A.}=\beta_{S}$, ${\rm Dec.}=
\delta_{S}$), the impact parameter can be estimated as
\begin{eqnarray}
b=r_{G}\sqrt{1-(\sin \delta_{S}\sin \delta_{G}+\cos \delta_{S}\cos \delta_{G}\cos (\beta_{S}-\beta_{G}))^2} \nonumber\\
\end{eqnarray}
where $r_{G}=8.3~{\rm kpc}$ is the distance from the Sun to the galaxy center, and ($\beta_{G}
={\rm 17^{h}45^{m}40.04^{s}}$, $\delta_{G}=-29^{\circ}00'28.1''$) are the coordinates of the galaxy 
center in the equatorial coordinate system \cite[J2000;][]{gillessen09}. Note that a more accurate formulation 
for $U_{\rm mw}(r)$ may provide higher order corrections to equation \ref{eq:gamma}, but the 
correction would be absorbed into the approximation obtained from replacing $U_{\rm IG}(r)$ and $U_{\rm 
host}(r)$ with the extension of $U_{\rm mw}(r)$. 

\section{EEP constraints with GRBs}

Gamma-ray bursts are the most extreme explosive events in the universe. They were initially observed 
as short, intense, and non-repeating flashes of $\sim {\rm MeV}$ $\gamma$-rays. Later on, it was found 
that the initial burst is usually followed by a longer-lived broadband emission (X-ray, ultraviolet, 
optical, infrared, microwave and radio), showing that GRBs are actually multi-wavelength transients, 
instead of simply bursts of $\gamma$-rays. The initial MeV $\gamma$-ray emission is often called the
``prompt emission”, while the longer wavelength emission is referred to as the ``afterglow emission”. 
It is worth noticing that in some bursts, the longer wavelength emission, e.g. optical emission, 
sometimes already starts during the prompt emission phase. The duration of the prompt emission 
is relatively short, ranging from $\sim0.1$ s to $\sim 1000$ s. Morever, the histogram of the 
duration shows a bimodal structure separated at 2 s, where  bursts with duration less than 2 s are 
classified as short-GRBs and those that last for more than 2 s are called long- GRBs. \cite[][for a 
review]{kumarzhang14}. 

GRBs are interesting cosmic transients for obtaining EEP constraints for the following reasons:
\begin{itemize}
\item Even the $\gamma$-ray photons are usually observed in different energy bands, for instance, 
the two instruments onboard Fermi, the Gamma-ray Burst Monitor (GBM; \cite{meegan09}) and the 
Large Area Telescope (LAT; \cite{atwood09}), provide unprecedented spectral coverage for seven 
orders of magnitude in energy (from $\sim$8 keV to $\sim$300 GeV). 
\item The observed $\gamma$-ray lightcurves in each energy band often have violent variability, 
with minimum timescales as small as milliseconds \citep{bhat92}. It is easy to use these sharp 
features to identify the arrival time lag between different energy photons. 
\item Thanks to the prompt slewing capability of the X-Ray Telescope (XRT; \cite{burrows05}) and the
UV-Optical Telescope (UVOT; \cite{roming05}) onboard the Swift satellite \citep{gehrels04}, longer 
wavelength photons have been observed at very early times after the GRB trigger, even during the 
prompt emission phase. It is thus possible to determine the arrival lag between $\gamma$-ray photons 
and low energy photons, such as optical photons. 
\end{itemize}

After four decades of observation, thousands of GRBs have been studied in significant detail,
making it possible to find some good examples that can be used to constrain the EEP. 
In this work, we take two examples, GRB 090510 and GRB 080319B. 

\subsection{GRB 090510}

GRB 090510 first triggered the Fermi-GBM on a precursor at $T_0=00:22:59.97$ UT, May 10th, 2009, 
and the main emission episode in the $\rm 8~keV-40~MeV$ energy range started $\sim0.5$ s 
after $T_0$, lasting up to $\sim1$ s. Then, $0.43$ s after the GBM trigger, it was detected and
precisely localized by the \textit{Swift} satellite, with coordinates (J2000) ${\rm R.A.= 22^{h}14^{m}
12.47^{s}}$, $\rm Dec.=-26^{\circ}35'00.4''$ \citep{hoversten09}. The emission observed by the Fermi-
LAT started 0.65 s after the GBM trigger and lasted $\sim200$ s \citep{guiriec09,ohno09}. The 
$T_{90}$ of the burst is $0.3$ s, placing it in the short GRB category. Subsequent observations 
determined its redshift as $z=0.903\pm0.003$ \citep{rau09,mcbreen10}. 

\cite{abdo09} analyzed in detail its prompt light curve over the different energy bands observed by 
Fermi. They found that the total GBM light curve ($\rm 8~keV-40~MeV$) consists of seven main pulses, 
where the first one is a dim short spike. A single 31-GeV photon 
was detected by LAT at 0.83 s after the GBM trigger, which coincides in time with the last of 
the seven GBM pulses. The directional and temporal coincidence of this photon with the MeV radiation
of GRB 090510 is very significant, at the $>5\sigma$ confidence level \citep{abdo09}. No high-energy 
(GeV) photon has ever been detected before the onset of the low-energy (MeV) emission in a GRB, and
this 31-GeV photon similarly appears likely to have been emitted after the onset of GRB 090510. 
Taking as our nominal time delay $\Delta t \simeq 0.83$ s, we thus have from eq.(\ref{eq:gamma}) 
for this burst
\begin{eqnarray}
\gamma_{\rm GeV}-\gamma_{\rm MeV}<2\times 10^{-8}
\end{eqnarray}

\subsection{GRB 080319B}

GRB 080319B was detected by Swift at $T_0=06:12:49$ UT, March 19th, 2008, with coordinates 
(J2000) ${\rm R.A.= 14^{h}31^{m}40.7^{s}}$, $\rm Dec.=+36^{\circ}18'14.7''$ \citep{racusin08a}. It 
is a long burst with $T_{90}>50$s and located at $z=0.937$ \citep{vreeswijk08}. At the trigger time of 
this burst, the wide-field robotic optical telescope ``Pi of the Sky" \citep{cwiok07} and the wide-
field robotic instrument Telescopio Ottimizzato per la Ricerca dei Transienti Ottici Rapidi (TORTORA; 
\cite{molinari06}) both serendipitously had the GRB within their fields of view  (as they were both 
already observing GRB 080319A). A bright optical transient was observed starting at $2.75\pm 5$s 
and peaking around 18s after the Swift trigger. The peak V-band magnitude reached 5.3, 
corresponding to a flux density of $\sim$28.7 Jy \citep{karpov08,zou09}. TORTORA measured the 
brightest portion of the optical flash with high time resolution, distinguishing three separate 
peaks which enabled doing detailed comparisons between the prompt optical and $\gamma$-ray 
emissions. \cite{kumar09} calculated the optical-$\gamma$-ray correlation function for the entire 
prompt emission, and they found that for the contemporaneous rise at $0-18$ s and fall at $43-60$ 
s, the correlation coefficient has a maximum at lags between 0 and 5 s , while for the middle part of 
the prompt emission, the correlation function still has a peak at lags between $-1$ to 3 s, but is less 
prominent. To be conservative, we adopt the largest value $5$ s as the arrival lag $\Delta t$ between 
the MeV photons and the optical photons. From eq.(\ref{eq:gamma}) we thus have for this burst
\begin{eqnarray}
\gamma_{\rm eV}-\gamma_{\rm MeV}<1.2\times 10^{-7}
\end{eqnarray}

\section{Discussion}

In the last two sections, we have proposed a practical method for using cosmic transients for
testing the energy dependence of the $\gamma$ parameter quantifying departures from the EEP,
and we have taken GRBs as an example to illustrate the capabilities of this method. In the following, 
we discuss some of the uncertainties involved in this method and the caveats on our constraint results.

The largest uncertainty is from the intrinsic time delay between two test photons, $\Delta t_{\rm int}
$. Since we are only seeking upper limits, in the calculations we neglected the $\Delta t_{\rm int}$ 
term based on the assumption that $\Delta t_{\rm int}$ has the same sign as $\Delta t_{\rm obs}$. More
specifically, here we assumed that the GeV (optical) photons are intrinsically emitted later than 
(or at most simultaneously with) the MeV photons in GRB 090510 (GRB 080319B). From the observational 
point of view, such an assumption is empirically justified, since we have never observed any GRBs with 
GeV or optical photons arriving earlier than the MeV photons. In standard GRB models, the MeV photons 
are assumed to arise in internal shock regions via synchrotron emission, while GeV photons are usually 
believed to originate from the inverse Compton up-scattering of MeV photons, and the optical photons 
are usually assumed to arise in the external shock region (which is outer and occurs later than the 
internal shock) also via synchrotron emission. In other words, from the theoretical point of view, our 
assumption about the intrinsic order of GeV or optical photons with MeV photons is also standard
practice. However, if in reality the GeV or optical photons are emitted earlier than the MeV photons, 
and $|\Delta t_{\rm int}|\gg|\Delta t_{\rm obs}|$, our constraint results would be invalidated.

The second uncertainty comes from the determination of $\Delta t_{\rm obs}$. For GRB 090510, we 
chose $\Delta t_{\rm obs}$ as the time delay between the 31-GeV photon and the trigger photons 
for the Swift satellite. Our results could be more stringent if in reality the 31-GeV photon were
correlated with the other GBM (MeV) pulses, for example the seventh pulse which coincides in time with 
the GeV photon \citep{abdo09}. However, in principle, it would be conceivable that the GeV photon is 
correlated with some other MeV photons emitted earlier but which failed to trigger the Swift satellite. 
This would be similar to a case where intrinsically this GeV photon is emitted earlier than the MeV 
photons, which could potentially invalidate our results. For GRB 080319B, we adopted the time delay 
results from the optical-$\gamma$-ray correlation function. These results should be reliable, since 
the TORTORA telescope provided high time resolution measurements for the brightest portion of the 
optical flash. Furthermore, it is worth noting that our adopted delay time is even larger than the 
time delay between the onset times of the optical observations and the MeV trigger. 

The total mass and the exact gravitational potential function of our galaxy is not well known. More 
accurate values for $M_{\rm mw}$ or for the function $U_{\rm mw}(r)$ could improve our constraints, 
but the correction should be limited to less than an order of magnitude.

In the calculations presented we neglect the time delay terms caused by the host galaxy and the 
intergalactic background gravitational potential. In principle, these terms could be much larger than 
$\Delta t_{\rm gra,mw}$. With future observations, and with a better understanding of the potential 
functions for these terms, our result could be improved by orders of magnitude.

\section{Conclusion}

In principle, the accuracy of the EEP may be characterized by limits on the differences between the
$\gamma$ values (one of the PPN parameters) for different species of particles, or for the same specie 
of particle at different energies. In this work, we propose that the time delays between correlated 
photons from cosmological transients, such as GRBs, can be used to constrain the accuracy of the EEP. 
Using the data from two observed GRBs as an example, and assuming that the observed time delays 
between photons of different energies are caused dominantly by the gravitational effects of our galaxy, 
an assumption which we justified in \S 2, for the GeV and MeV photons of one burst we give  an upper limit
of $\gamma_{\rm GeV}-\gamma_{\rm MeV}<2\times 10^{-8}$; and for the optical and MeV photons form 
another burst  we find $\gamma_{\rm eV}-\gamma_{\rm MeV}<1.2\times 10^{-7}$. 

Previous to this present work,
the most accurate constraints on $\gamma$ differences were based on the photons and neutrinos 
from Supernova 1987A, giving an upper limit of $0.37\%$ for optical photons and MeV neutrinos, 
and an upper limit of $1.6\times 10^{-6}$ for two neutrinos, whose energy difference was 
of order $\sim$ MeV. If our present interpretation is correct, our results increase the accuracy 
level of such EEP constraints by at least one order of magnitude, to $\sim10^{-7}$, while 
extending the energy range tested to the GeV-MeV and MeV-eV range.

A new type of transients, called  Fast Radio Bursts, have lately attracted attention \citep{thornton13}. Although their physical origin is still debated \citep{kulkarni14}, if in the future they are proven to be cosmic transients, their simple sharp features make them attractive candidates for constraining the EEP, which could further extend the tested energy range to the radio band with high accuracy.

It is worth pointing out that high energy neutrinos have long been proposed to be associated with 
GRBs, although a dedicated search of high-energy neutrinos coincident with electromagnetically detected
GRBs has so far led to null results \citep{abbasi12,aartsen15}. If in the future GRBs with both high 
energy photons and neutrinos are detected, better constraints on EEP might be achieved. On the other 
hand, neutrino detectors such as IceCube have reached the sensitivity to detect high-energy neutrinos
of astrophysical origin for the first time. The three-year discovery catalog of IceCube contains a total 
of 37 neutrino candidate events with deposited energies ranging from 30 to 2000 TeV \citep{aartsen14}. 
Although the origin of these neutrinos is still debated, some of them seem to have both spatial and 
temporal correlations, such as events 24 and 25 in the catalog \citep{aartsen14}. In the future, if the 
origin of these neutrinos is better understood, or the association between different detected events 
is confirmed, it will be essential to use these high energy cosmic neutrinos for obtaining better 
constraints on the accuracy of the EEP.

The most precise previous constraints on the absolute $\gamma$ value of photons is 
$\gamma-1\sim10^{-5}$, which comes from radio photons. Our relative limits are three orders 
of magnitude lower than this,  focusing on the value of the difference between various energies, 
rather than the absolute value. On the other hand, the Shapiro time-delay measurements using the 
Hipparcos optical astrometry satellite yielded an agreement with General Relativity of 
$\gamma_{eV} -1 \lesssim 0.3\%$ \citep{froeschle97}. Based on our results, we can predict that 
the constraint on the absolute $\gamma$ values MeV or GeV photons, which from the sources 
we investigated here differ from that of their corresponding optical photons by 
$\lesssim 10^{-7}$, should agree with GR at least to this same $\lesssim 0.3\%$ level of accuracy.
Thus, we have extended the validity of this absolute value bound on $\gamma$ at this level 
from the optical to the GeV range, and have constrained the relative variation of $\gamma$ between 
optical and GeV to the $\lesssim 10^{-7}$ range.

\acknowledgments

We thank the anonymous referee for a valuable report and Derek Fox, Bing Zhang, Zi-Gao Dai, Xiang-Yu Wang and Mou-Yuan Sun for helpful comments. This work is supported by National Basic Research Program ('973' Program) of China (grants 2014CB845800 and 2013CB834900), NASA NNX 13AH50G, the National Natural Science Foundation of China (grants nos. 11322328 and 11433009). XFW is partially supported by the One-Hundred-Talents 
Program, the Youth Innovation Promotion Association, and the Strategic Priority Research Program 
``The Emergence of Cosmological Structures'' (grant no. XDB09000000) of of the Chinese Academy 
of Sciences, and the Natural Science Foundation of Jiangsu Province (grant no. BK2012890).

\end{document}